\documentclass[twocolumn,showpacs]{revtex4}

\usepackage{graphicx}
\usepackage{dcolumn}
\usepackage{amsmath}
\makeatletter
\def\btt#1{\texttt{\@backslashchar#1}}
\DeclareRobustCommand\bblash{\btt{\@backslashchar}}
\makeatother

\begin{document}

\title{Exact non-spherical relativistic star}

\author{S. G.~Ghosh}
\email{sgghosh@iucaa.ernet.in} \affiliation{BITS, Pilani DUBAI
Campus, P.B. 500022, Knowledge Village, DUBAI, UAE  and
Inter-University Center for Astronomy and Astrophysics,
 Post Bag 4 Ganeshkhind,  Pune - 411 007, INDIA }
\author{D. W.~Deshkar}
\affiliation{S.S.E.S. Amti's Science College, Congress Nagar,
Nagpur-440 012, India}

\date{\today}

\begin{abstract}
We obtain Vaidya-like solutions to include both a null fluid and a
string fluid in non-spherical (plane symmetric and cylindrical
symmetric) anti-de Sitter space-times. Assuming that string fluid
diffuse, we find exact solutions of Einstein's field equations. Thus
we extend a recent work of Glass and Krisch \cite {gk} to
non-spherical anti-de Sitter space-times.
\end{abstract}


\maketitle

\section{Introduction}
The string is very important ingredient in many physical theories
and idea of string is fundamental to superstring theories \cite{as}.
The development of the last few years have opened the possibility to
go beyond perturbation theory and to address the most interesting
problems of quantum gravity. The intense level of activity in string
theory has lead to the idea that many of the classical vacuum
scenario may have atmosphere composed of a fluid or strings.
Recently, Glass and Krisch \cite{gk} extended the classical Vaidya
solution \cite{pc} to create two fluid atmosphere outside the star,
the original null fluid and a new fluid composed of string, and we
shall call it generalized Vaidya solution. The solution is very
important in view of recent links between black holes and string
theories \cite{mc,sk,fl}.

This solution has been employed to look into the consequence of
string fluid on the formation of naked singularities in Vaidya
collapse \cite{gg,gs}. Indeed, the effect of string is a shrinkage
of the naked singularity initial data space, or an enlargement of
the black hole initial data space of Vaidya collapse \cite{gs} and
thus violating {\em cosmic censorship conjecture} (CCC) \cite{rp}.
However, the conjecture, in it's weak version,  essentially state
that any naked singularity which is created by evolution of regular
initial data will be shielded from the external view by an event
horizon. According to the strong version of the CCC, naked
singularities are never produced, which in the precise mathematical
terms demands that space-time should be globally hyperbolic.  Many
of the gravitational collapse studies were motivated by Thorne's
hoop conjecture \cite{kt1} that collapse will yield a black hole
only if a mass $M$ is compressed to a region with circumference $C
\leq 4 \pi M$ in all directions. If hoop conjecture is true, naked
singularities may form if collapse can yield  $C \geq 4 \pi M$ in
some direction.  Most of the studies in collapse were restricted to
spherically symmetric space-times.

 On the other-hand, non-spherical collapse not
so well understood. However, non-spherical collapse could occur in
real astrophysical situation, and it is also important for a better
understanding of both cosmic censorship conjecture and hoop
conjecture. Also, hoop conjecture was given for space-times with a
zero cosmological term and also Penrose \cite{rp1} has conjectured
that CCC might require a zero cosmological term. In the presence of
negative cosmological term one can expect the occurrence of major
changes. Indeed, Lemos \cite{jl} has shown that planar or
cylindrical black holes form rather than naked singularity from
gravitational collapse of a planar or cylindrical matter
distribution in an anti-de Sitter space-time, violating in this way
the hoop conjecture but not CCC. Cai {\it et al.} \cite{cz} extended
the work of Lemos \cite{jpl} to the plane symmetric and
cylindrically symmetric solutions in Einstein-Maxwell equations with
a negative cosmological term and pointed out that the negative
cosmological term plays a crucial role in their solutions, as in the
BTZ black holes.

 The purpose of this brief report
is to see how the results that were presented in \cite{gk} get
modified for non-spherical (plane symmetric and cylindrical
symmetric) space-times with a negative cosmological constant. We are
able to provide a two-fluid kinematic interpretation of the
transverse stresses, and the null-tetrad and the orthogonal vectors
for the non-spherical space-times.  Exact solutions of the Einstein
field equations are obtained assuming that the string fluid diffuse.

\section{Non-spherical two fluid model}
In this section, we extend the work of the Glass and Krisch
\cite{gk} to non-spherical (plane symmetric and cylindrical
symmetric) anti-de Sitter space-times. Let us first consider the
case of plane symmetry. The metric of general plane symmetric
space-time, expressed in terms of Eddington retarded time coordinate
(ingoing coordinate) $u$, reads:
\begin{equation}
ds^2 = e^{\psi(u,r)}du \left[e^{\psi(u,r)} f(u,r) du
 -  2 dr \right] - r^2 \alpha^2 (dx^2+ dy^2), \label{eq:me}
\end{equation}
where  $- \infty \leq x, \: y \leq \infty$ are coordinates which
describe two-dimensional zero-curvature space which has topology $R
\times R$,  $- \infty \leq u  \leq \infty$ is null coordinate called
the retarded time, and  $0  \leq r \leq \infty$ is the radial
coordinate. Further, $e^{\psi(u,r)}$ is an arbitrary function and
where $3 \alpha^2 = - \Lambda > 0$ denote negative cosmological
constant. It is useful to introduce a local mass function $m(u,r)$
defined by $f = 1 - {2 m(u,r)}/{r}$. For $f = m(u)/r$ and $\psi=0$,
the metric (\ref{eq:me}) reduces to the \emph{plane symmetric
Vaidya-like metric} \cite{bk,cs}. Initially $f= M_0/r$ (with
$\psi=0$) provides the vacuum Taub solution \cite{at}.

It is the field equation $G^0_1 = 0$ that leads to $ e^{\psi(u,r)} =
g(u)$. However, by introducing another null coordinate $
e^{\psi(u,r)} = g(u)$, we can always set without the loss of
generality, $\psi(u,r) =0$. Hence, the metric takes the form
\cite{cz,sg}:
\begin{equation}
ds^2 =  \left[1 - \frac{2 m(u,r)}{r}\right] d u^2 - 2 d u d r -
\alpha^2 r^2 (dx^2+ dy^2). \label{eq:me1}
\end{equation}
The use of a Newman-Penrose null tetrad formalism leads to Einstein
tensor of the form \cite{gk,mc}:
\begin{equation}
G_{ab} = - 2 \Psi_{11}(l_a n_b + l_b n_a + m_a \overline{m}_b +
\overline{m}_a m_b) - 2 \Psi_{11} l_a n_b - 6\Lambda. \label{ee}
\end{equation}
Here the null tetrad Ricci scalars are
\begin{eqnarray}
  \Psi_{11} = \frac{1}{r^2} \left[{2 \frac{\partial m}{\partial r} - r \frac{\partial^2 m}{\partial r^2} -
  1}\right], \\
  \Psi_{22} = - \frac{1}{r^2} \frac{\partial m}{\partial v} \\
  \Lambda = \frac{1}{r^2} \left[{2 r \frac{\partial^2 m}{\partial r^2} - 2 \frac{\partial m}{\partial r} -
  1}\right],
\end{eqnarray} the
principal null geodesic vectors are $l_a,n_a$ of the form
\begin{equation}
l_a=\delta _a^u,\qquad n_a=f/2\delta _a^u+\delta _a^r,
\end{equation}
where $l_al^a=n_an^a=0,$ $l_an^a=-1$.  The metric (\ref{eq:me1})
admits an orthonormal basis defined by four unit vectors
\begin{eqnarray}
\hat{u}_a=f^{1/2}\delta _a^u+f^{-1/2}\delta _a^r,\qquad \hat{r}%
_a=f^{-1/2}\delta _a^r, \\
\hat{\mathbf{x}}_a=\alpha x\delta _a^x ,\qquad \hat{%
\mathbf{y}}_a=\alpha y \delta _a^y ,  \label{ov}
\end{eqnarray}
where $\hat{u_a}$ is a timelike unit vector and $\hat{r_a}$,
$\hat{\mathbf{x}_a}$, $\hat{\mathbf{y}_b}$ are unit space-like
vector such that
\begin{equation}
g_{ab} = \hat{u_a}\hat{u_b} - \hat{r_a}\hat{r_b} -
\hat{\mathbf{x}_a}\hat{\mathbf{x}_b}- \hat{\mathbf{y}_a}
\hat{\mathbf{y}_b}. \label{mt1}
\end{equation}
Associated with the string worldsheet we have the string bivector
defined by
\begin{equation}
\Sigma^{ab}=\epsilon^{AB} \frac{dx^{a}}{dk^A}\frac{dx^{b}}{dk^B},
\label{sb}
\end{equation}
where $\epsilon^{AB}$ is two dimensional Levi-Civita symbol.  It is
useful to write the bivector, in terms of the unit vectors, as
\begin{equation}
\Sigma ^{ab}=\hat{r}^a\hat{u}^b-\hat{u}^a\hat{r}^b,  \label{bv}
\end{equation}
and the condition that the worldsheet are timelike, i.e., $\gamma
=1/2\Sigma ^{ab}\Sigma _{ab}<0$ implies that only the $\Sigma ^{ur}$
component is non-zero, therefore one obtains:
\begin{equation}
\Sigma ^{ac}\Sigma _c^b=\hat{u}^a\hat{u}^b-\hat{r}^a\hat{r}^b.
\end{equation}
The string energy-momentum tensor for a cloud of string, by analogy
with the one for the perfect fluid, is written as \cite{gk,mc}:
\begin{equation}
T_{ab}^{\left( s\right) }=\rho \Sigma _a^c\Sigma _{cb}-p_{\bot
}h_{ab}s \label{ems}
\end{equation}
The energy-momentum of two fluid system is $T_{ab} = T^{(n)}_{ab} +
T^{(s)}_{ab}$, where
\begin{equation}
T_{ab}^{(n)}=\psi l_al_b.
\end{equation}
It is the null fluid tensor corresponding to the component of the
matter field that moves along the null hypersurfaces
$u=\mbox{const.}$ The effective energy momentum tensor for two fluid
system, in terms of the unit vectors, can be cast as:
\begin{equation}
T_{ab}=\psi l_al_b + \rho
\hat{u}_a\hat{u}_b+p_r\hat{r}_a\hat{r}_b+p_{\bot }\left(
\hat{\mathbf{x}}_a\hat{\mathbf{x}}_b+\hat{\mathbf{y}}_a\hat{\mathbf{y}}_b\right).
\label{emsv}
\end{equation}
For $\rho = p_r = p_{\bot }= 0$, Eq.~(\ref{emsv}) reduces to
stress-energy tensor which gives Vaidya metric \cite{bk,cs}. The
Einstein field equations ($G^a_b - 3 \alpha^2 \delta^a_b = T^a_b$)
now take the form:
\begin{eqnarray}
\psi =\frac{1}{4 \pi r^2} \frac{\partial m}{\partial u}, \label{psi}\\
\rho =-p_r= - \frac{1}{8\pi r^2}+\frac{1}{4\pi r^2}\frac{\partial
m}{\partial r} + \frac{1}{8\pi } 3 \alpha^2, \label{rho}\\
p_{\bot }= - \frac{1}{8\pi r} \frac{\partial^2 m}{\partial r^2} -
\frac{1}{8\pi } 3 \alpha^2.\label{tp}
\end{eqnarray}

\subsection{Diffusive mass solutions} A typical approach to characterizing diffusive
transport of the particles begins with Fick's law of diffusion.  The
Fick's law, widely used in transport theory, is a phenomenological
statement of a macroscopic nature about the relationship between the
mass current $\vec{J}_{n}$ and the particle density $n$ of a fluid.
More precisely put,
\begin{equation}
\vec{J}_{(n)} = -\mathcal{D} \vec{\nabla}n,
\end{equation}
introducing the diffusion constant $\mathcal{D}$, called Fick's
diffusion constant and where $\vec{\nabla}$ is a purely spatial
gradient. Then 4-current conservation $J^{\mu}_{(n);{a}} = 0$, where
\begin{equation}
J^{a}_{(n)}\partial_{a} = (n,\vec{J}_{(n)})\nonumber\\
= n \partial_{u} - \mathcal{D}\left(\frac{\partial n}{\partial
r}\right)\partial_{r}.
\end{equation}
If Fick's law is combined with the continuity equation, there
results the diffusion equation
\begin{equation}
\partial_{u} n = \mathcal{D}\nabla^2 n.\label{de}
\end{equation}
Here we assume that the string diffuse and that diffusion is the
movement of particle form higher number to lower number. By
rewriting the $T_{a b}$ components as
\begin{eqnarray}
\frac{\partial m}{\partial u} = 4\pi r^2 \psi,\label{mu}\\
\frac{\partial m}{\partial r} = 4\pi r^2 \rho - \frac{3}{2} \alpha^2
r^2 + \frac{1}{2}, \label{mr}
\end{eqnarray}
The integrability condition,
\begin{equation}
\frac{\partial}{\partial r}\left(\frac{\partial m}{\partial
u}\right) = \frac{\partial}{\partial u}\left(\frac{\partial
m}{\partial r}\right),
\end{equation}
with the help of Eqs.~(\ref{mu}) and (\ref{mr}), takes the form
\begin{equation}
\frac{\partial \rho}{\partial u} = r^{-2} \partial_{r}(r^2 \psi).
\label{der}
\end{equation}
Rewriting the diffusion equation (\ref{de}) ($n$ replaced by $\rho$)
\begin{equation}
\frac{\partial \rho}{\partial u} = \mathcal{D} r^{-2}
\frac{\partial}{\partial r}(r^2 \frac{\partial \rho}{\partial r}),
\label{der1}
\end{equation}
and comparing with ${\partial \rho}/{\partial u}$ in
Eq.~(\ref{der}),we obtain
\begin{equation}
\frac{\partial m}{\partial u} = 4 \pi \mathcal{D} r^2 \frac{\partial
\rho}{\partial r}.\label{mu1}
\end{equation}
Here we have omitted the function of integration, because it can be
set to zero by suitable redefinition of coordinate. Thus solving the
diffusion equation for $\rho$ and then integrating only those
solutions to obtain $m$ that provides exact Einstein solutions for
diffusing string fluids. There are many analytic solutions of
Eq.~(\ref{der1}) and three of them are
\begin{eqnarray}
\rho = \rho_{0} + \frac{k_{1}}{r},\label{s1}\\
 \rho = \rho_{0} +
(k_{2}/6) r^2 + k_{2} \mathcal{D}u, \label{s2} \\
\rho = \rho_{0} + k_{3}(\mathcal{D}u)^{-3/2} \exp\left[\frac{-r^2}{4
\mathcal{D} u} \right].\label{s3}
\end{eqnarray}
Where $\rho_{0}$ is the density at spatial infinity and $ k_{1},
k_{2}, k_{3}$ are constants. The physical behavior of the density
solutions provides a variety of atmospheric models. The density
solutions are identical to the analogous spherically symmetric
solutions \cite{gk}.  Next, integrating Eq.~(\ref{mr}) and
Eq.~(\ref{mu1}), we obtain the following mass equation:
\begin{equation}
m(u, r) = m_{0} + 4 \pi \int r^2 \rho dr + 4 \pi r^2 \mathcal{D}\int
\frac{\partial \rho}{\partial r} du - \frac{1}{2}\alpha^2 r^3 +
\frac{1}{2}r. \label{mass}
\end{equation}
The corresponding masses with the density solutions above are:
\begin{widetext}
\begin{eqnarray}
m(u, r) = m_{0} + \frac{4}{3} \pi r^3 \rho_{0} + 2 \pi k_{1}(r^2 - 2
\mathcal{D}u) - \frac{1}{2}\alpha^2 r^3 +
\frac{1}{2}r, \label{m1}\\
m(u, r) = m_{0} + \frac{4}{3} \pi r^3 \rho_{0} + \frac{4}{3} \pi
k_{2}r^2\mathcal{D}u (r + 1)+ \frac{2}{15} \pi k_{2}r^5 -
\frac{1}{2}\alpha^2 r^3 +
\frac{1}{2}r, \label{m2}\\
m(u, r) = m_{0} + \frac{4}{3} \pi r^3 \rho_{0} + \frac{2}{3} \pi
k_{3} r^3(\mathcal{D}u)^{-3/2}\exp \left[-\frac{r^2}{4 \mathcal{D}
u}\right] - \frac{1}{2}\alpha^2 r^3 + \frac{1}{2}r. \label{m3}
\end{eqnarray}
\end{widetext}
Here The family of solutions discussed above, in general, belongs to
Type II fluid defined in \cite{he}.  However, it is interesting to
note that if parameters $k_{1}, k_{2}$ and $k_{3}$ are set to zero
then all the mass solutions above become the static solution:
\begin{equation}
m(r)= m_{0} + \frac{4}{3} \pi r^3 \rho_{0} - \frac{1}{2}\alpha^2 r^3
+ \frac{1}{2}r.\label{mu2}
\end{equation}
Substituting this mass function Eq.~(\ref{mu2}), in
Eqs.~(\ref{psi}), (\ref{rho}) and (\ref{tp}), yields
\begin{eqnarray}
\psi = 0,\\
\rho = -p_{r} = \rho_{0},\\
p_{\bot} = -\rho_{0}.
\end{eqnarray}
Thus, we have $m=m(r)$ and $\psi$=0, so the matter field degenerates
to type I isotropic string fluid \cite{ww}. \\

We turn our attention to the cylindrical symmetric ant-de Sitter
space-time. The metric (\ref{eq:me1}) in the Cylindrical space-time
has the form \cite{sg}:
\begin{equation}
ds^{2} = - \left[1 - \frac{2 m(u, r)}{r}\right] du^{2} + 2 du dr +
r^{2} d\theta^{2} + \alpha^{2} r^{2} dz^{2},
\end{equation}
where $-\infty < u, z < \infty$, $0 \leq r < \infty$ and $0 \leq
\theta \leq 2 \pi$. The topology of two dimensional zero curvature
space-time is $R \times S^{1}$.  The analysis given above, with
suitable modifications, is valid in  the cylindrical anti-de Sitter
space-time as well. Hence, to conserve space, we avoid repetition of
the detailed analysis (being similar to plane symmetric case).

\subsection{Isotropic string fluid}
The pressure isotropy implies that the radial pressure equal with
the transverse stress $p_r$ $=p_{\bot }$, then we have,
\begin{equation}
m(u,r) = \bar{ M}(u) + \frac{r^3}{3} \bar{ S}(u) + \frac{r}{2}
\end{equation}
Here,  $\bar{ M}(u), \bar{ S}(u)$ are arbitrary function, and as
above $ \bar{ M}(u)$ can be treated as Vaidya-mass and other term as
contribution from string fluid. Thus in this case mass function has
no explicit contribution form cosmological term.  The physical
quantities for this metric as in are given by
\begin{eqnarray}
\psi = \frac{1}{4\pi r^2} \left[\frac{\bar{{M}}(u)}{\partial
u}+\frac{\bar{{S}}(u)}{\partial u} r^3 \right], \nonumber \\
\qquad - \rho = p_r = p_{\bot } = p = - \frac{1}{4\pi} S(u) -
\frac{1}{8 \pi} 3 \alpha^2 \label{eq:rho1}
\end{eqnarray}
If we consider static mass function $m(r)$, then $p_r$ $=p_{\bot }$,
gives corresponding static solution:
\begin{equation}
m(r) = c_1 + c_2 \frac{r^3}{3}  + \frac{r}{2}
\end{equation}
Thus, we have $m=m(r)$ and $\psi$=0, so again the matter field
degenerates to type I isotropic string fluid \cite{ww}.
\subsection{Separable solutions}
Next, we see how spherically symmetric solutions of Govender and
Govender \cite{gg} gets modified in our non-spherical space-times.
Let us assume that
\begin{equation}
\rho = R(u)U(u),
\end{equation}
then, it is easy to see that:
\begin{equation}
\rho = e^{u \lambda}\left(\frac{C_{1}}{e^{r
\sqrt{\frac{\lambda}{\mathcal{D}}}} r} + \frac{C_{2} e^{r
\sqrt{\frac{\lambda}{\mathcal{D}}}}}{2 r
\sqrt{\frac{\lambda}{\mathcal{D}}}}\right),
\end{equation}
where $C_{1}$ and $C_{2}$ are constants and $\lambda$ is the
separation constant. This leads to a slightly complicated
expressions for the mass function:
\begin{eqnarray}
m(u, r) = m_{0} + \frac{4}{3} \pi r^{2} e^{\lambda u}
\frac{C_{1}}{e^{r \sqrt{\frac{\lambda}{\mathcal{D}}}}}\zeta_{-}(u) \nonumber\\
+ \frac{4}{3} \pi r^{2} \frac{C_{2}}{2
\sqrt{\frac{\lambda}{\mathcal{D}}}} e^{\lambda u} e^{r
\sqrt{\frac{\lambda}{\mathcal{D}}}} \zeta_{+}(u) -\frac{1}{2}
\alpha^{2} r^{3} + \frac{1}{2} r.
\end{eqnarray}
where, \[ \zeta_{\pm}(u)= \left(1 \pm \frac{3 \mathcal{D} u}{r}
\sqrt{\frac{\lambda}{\mathcal{D}}} - \frac{3 \mathcal{D} u}{r^{2}}
\right). \]

On the other hand, if we require that $\rho = R(r) + U(u)$, then
\begin{equation}
\rho = C_{0} + \lambda u + \frac{C_{1}}{r} + \frac{\lambda r^{2}}{6
\mathcal{D}},
\end{equation}
with corresponding mass function
\begin{eqnarray}
m(u, r) = m_{0} + \frac{4}{3} \pi r^{3} \left(C_{0} + 2 \lambda u
 \right) + 2 \pi C_{1} r^{2} +\nonumber\\
\frac{2}{15} \pi \frac{\lambda r^{5}}{\mathcal{D}}  - 4 \pi C_{1}
\mathcal{D} u - \frac{1}{2} \alpha^{2} r^{3} + \frac{1}{2} r.
\end{eqnarray}
Here, $M(u) = -4 \pi C_{1}\mathcal{D} u$ can be considered as Vaidya
mass and other terms in the above equation have contribution from
the string fluid and the cosmological term.
\subsection{Constant density string fluid ($\rho = \rho_0$ )}
Although strictly constant density is not completely realistic, the
particular analytic solution to Einstein field equation with this
equation of state has provided some insights concerning stars in
general relativity.  For example, the star represented by this
solution has property that it cannot remain in equilibrium if the
size-to-mass ratio is less than 9/4 \cite{ck}. To obtain the
constant density solution, we take $\rho = \rho_0$. With this it can
bee seen that Eq.~(\ref{rho}) can be easily integrated to yield:
\begin{equation}
m(u,r) = \tilde{{M}}(u)+ \beta^2 r^3 + \frac{r}{2}
\end{equation}
where $\beta^2 = (8 \pi \rho_0 -  3 \alpha^2) /6$ is a constant.
Because $\rho$ is the constant, the physical parameter can also be
trivially evaluated to give
\begin{equation}
\psi = \frac{1}{4\pi r^2}  \frac{\tilde{{M}}(u)}{\partial u}
 \qquad \ p_r = p_{\bot} = - \rho_0,
\end{equation}
and in this case also we have isotropic string fluid.
\section{Concluding remarks}
In this work we have discussed Vaidya-like solution in non-spherical
(planar and cylindrical)  anti-de Sitter space-times for a two-fluid
system: a null fluid and a string fluid. By assuming that string
diffuse analytical solutions of Einstein's field equations have been
obtained. The Vaidya's radiating star metric is today commonly used
for various purposes: As a testing ground for various formulations
of the CCC. As an exterior solution for models of objects consisting
of heat-conducting matter. Recently, it has also proved to be useful
in the study of Hawking radiation, the process of black-hole
evaporation \cite{rp2}, and in the stochastic gravity program
\cite{hv}.  The solutions presented here can be useful to get
insights of the string effects in these physical process that too in
non-spherical space-time.

Since, comparatively a very few studies have been done in
non-spherical collapse, it should be very interesting use these to
study string effects in the final fate of collapse. This work in
under progress \cite{sg1}.

\end{document}